\ifx\pdfoutput\undefined\else
  \pdfoutput=1
\fi
\documentclass[prd,aps,twocolumn,preprintnumbers,showpacs,nofootinbib,superscriptaddress,notitlepage,floatfix]{revtex4-1}
\usepackage{amsmath,graphicx,amssymb}
\usepackage[utf8]{inputenc}
\usepackage{xcolor}
\usepackage{hyperref}

\allowdisplaybreaks

\newcommand{\nn}{\nonumber}

\begin{document}

\title{Revisiting Quark Confinement in the Proton through the Force on Quarks}

\author{Ji-Xin Yu}
\email{yujx18@lzu.edu.cn}
\affiliation{Frontiers Science Center for Rare Isotopes, and School of Nuclear Science and Technology, Lanzhou University, Lanzhou 730000, China}

\author{Ao-Sheng Xiong}
\email{xiongash21@lzu.edu.cn, corresponding author}
\affiliation{Frontiers Science Center for Rare Isotopes, and School of Nuclear Science and Technology, Lanzhou University, Lanzhou 730000, China}

\author{Ji Xu}
\email{xuji@lzu.edu.cn, corresponding author}
\affiliation{Frontiers Science Center for Rare Isotopes, and School of Nuclear Science and Technology, Lanzhou University, Lanzhou 730000, China}

\author{Fu-Sheng Yu}
\email{yufsh@lzu.edu.cn, corresponding author}
\affiliation{Frontiers Science Center for Rare Isotopes, and School of Nuclear Science and Technology, Lanzhou University, Lanzhou 730000, China}

\author{Yong Zheng}
\email{yzheng2018@lzu.edu.cn, corresponding author}
\affiliation{Frontiers Science Center for Rare Isotopes, and School of Nuclear Science and Technology, Lanzhou University, Lanzhou 730000, China}

\begin{abstract}
Quark confinement, the fact that colored quarks are permanently bound inside color-neutral hadrons and have never been observed as isolated particles, remains one of the central issues of the Standard Model. Recently, Ji et al.\,\cite{Ji:2026lyj} proposed a framework to define and measure the force on quarks in the proton, obtaining strong evidence for a net confining force and thus opening a new perspective on the study of confinement. In this work, we improve this analysis by incorporating light-cone QCD sum rule results to supplement the limited experimental and lattice QCD information in the large-$|q^2|$ region. We further formulate the reconstruction of the quark force as a regularized inverse problem, thereby reducing the model dependence associated with the prescribed functional parametrizations used before. The resulting quark force provides a complementary, less parametrization-dependent determination and remains consistent with that implied by a linear QCD potential, which also supports the robustness of the framework proposed in Ref.\,\cite{Ji:2026lyj}. We also show that improved future inputs can substantially reduce the uncertainty in the reconstructed quark force.
\end{abstract}

\maketitle

\section{Introduction}
\label{sec:introduction}

Quark confinement is one of the most fundamental and long-standing problems in particle physics. Although QCD has been firmly established as the theory of strong interactions, the mechanism by which colored quarks are permanently bound into color-neutral hadrons remains a central challenge. A complete understanding of confinement requires much more than just observing the absence of free quarks; it calls for a quantitative description of how the strong interaction generates binding forces on quarks inside hadrons.

Substantial progress has recently been made in understanding the mechanical properties of hadrons through pressure and shear force distributions, encoded mainly in the gravitational or energy-momentum tensor form factors (EMT FFs), which provide a compelling picture of how internal stresses stabilize the proton \cite{Ji:1996ek,Polyakov:2002yz,Polyakov:2018zvc,Shanahan:2018nnv,Tanaka:2018wea,Hagler:2003jd,Chen:2001pva,Gockeler:2003jfa}. These studies, including phenomenological extractions and lattice QCD calculations, have opened a promising direction for investigating quark confinement. However, the pressure and shear forces characterize the stress balance of the proton as a composite system. They do not directly give the color force acting on quarks at a given position inside the proton, and therefore are not the most direct quark-level measure of confinement. To address this limitation, Ji et al.\,\cite{Ji:2026lyj} proposed a direct way to define and measure the color-Lorentz force acting on quarks inside the proton through the divergence of the quark part of the EMT. In the infinite-momentum frame (IMF), this construction relates the transverse quark force to the quark scalar form factor $G_{s,q}(q^2)$. Their analysis provides compelling evidence for an attractive, approximately constant transverse force over an intermediate range of transverse distances, consistent with the expected behavior of confinement \cite{Bali:2000gf,Kawanai:2011xb,LealFerreira:1979xq}.

Despite this important advance, the available information on the relevant proton EMT FFs remains sparse at large spacelike momentum transfer. In particular, the analysis of Ref.\,\cite{Ji:2026lyj} used the GYZ results \cite{Guo:2025jiz}, which cover only the range $0\leq |q^2|\leq 2\,{\rm GeV}^2$, together with the GUMP and BEG results \cite{Guo:2025muf,Burkert:2018bqq}, which are restricted to $0\leq |q^2|\leq 0.35\,{\rm GeV}^2$. Consequently, the EMT inputs employed in that analysis provide little direct constraint in the large-$|q^2|$ region. This limited momentum-transfer coverage is particularly consequential for the Fourier-Bessel reconstruction of the transverse force distribution, which requires reliable knowledge of the form factors over a broad kinematic range. The scalar form factor entering this reconstruction is given by
\begin{eqnarray}\label{eq:Gsq_intro}
G_{s,q}(q^2)=A_q(q^2)+\frac{q^2}{4M^2}B_q(q^2)-\frac{3q^2}{M^2}C_q(q^2) \,,
\end{eqnarray}
so that the large-$|q^2|$ behavior of $A_q(q^2)$ and, in particular, $C_q(q^2)$ directly affects the spatial gradient that determines the force. Light-cone QCD sum rules (LCSRs) provide a standard nonperturbative framework for evaluating hadronic form factors at intermediate and large spacelike momentum transfer \cite{Balitsky:1989ry,Colangelo:2000dp,Azizi:2019ytx,Dehghan:2025ncw,Anikin:2019kwi}. For the proton, LCSR calculations provide complementary information on the quark EMT FFs over $1\lesssim |q^2|\lesssim 10\,{\rm GeV}^2$. This range extends substantially beyond the momentum-transfer coverage of the inputs used in Ref.\,\cite{Ji:2026lyj} and is therefore particularly valuable for reconstructing the force at short and intermediate transverse distances. We consequently use the LCSR results as complementary large-$|q^2|$ input for the quark EMT FFs entering $G_{s,q}(q^2)$, thereby improving the reconstruction of the transverse force on quarks in the proton.

\begin{figure}[ht]
    \centering
    \includegraphics[width=1.0\linewidth]{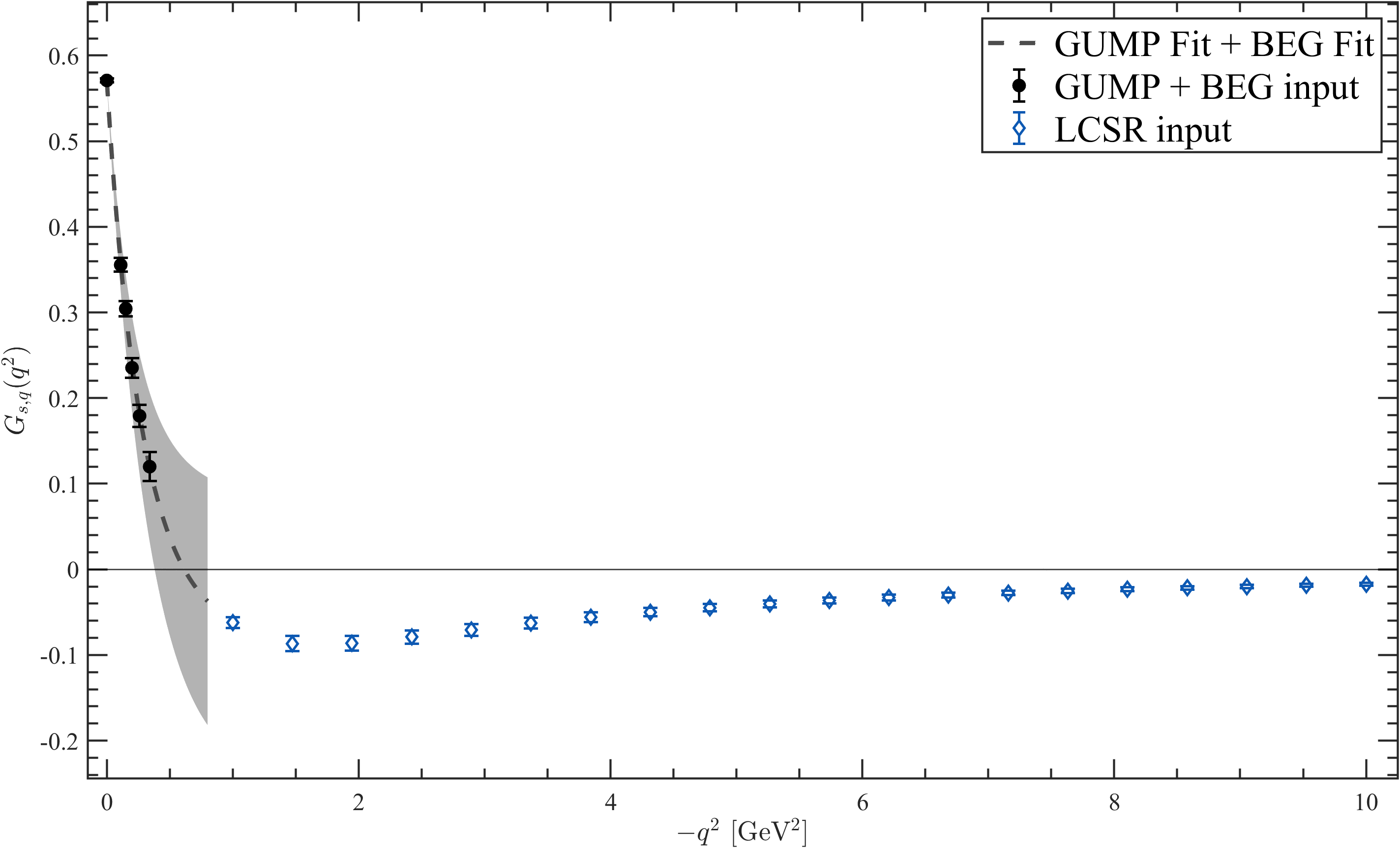}
    \caption{Input values of the quark scalar form factor $G_{s,q}(q^2)$ from the GUMP \cite{Guo:2025muf} and BEG \cite{Burkert:2018bqq} analyses (black points) and the LCSR calculations \cite{Dehghan:2025ncw} (blue points). A $10\%$ relative error bar is assigned to each point in the LCSR calculations.}
    \label{fig:input-points}
\end{figure}

A second source of uncertainty comes from the parametrization of EMT FFs. In Ref.\,\cite{Ji:2026lyj}, $G_{s,q}(q^2)$ was obtained through phenomenological parametrizations of $A_q(q^2)$ and $C_q(q^2)$, with the contribution of $B_q(q^2)$ neglected \cite{Burkert:2018bqq,Guo:2025muf}. While this strategy provides a practical first determination of the transverse force distribution, it inevitably introduces parametrization dependence, especially in the large-$|q^2|$ region. From a mathematical perspective, the available input consists of only a finite set of uncertain momentum space constraints, whereas the desired force is a continuous function in transverse coordinate space. The reconstruction must therefore bridge the incomplete momentum space information and the corresponding spatial distribution. Therefore, these limitations suggest that the determination of the quark force $\vec{F}_q(\vec{r}_\perp)$ is naturally an inverse problem: one seeks to reconstruct a continuous spatial distribution from finite and uncertain momentum space information \cite{Kirsch-2011,Li:2020xrz,Xiong:2022uwj,Xiong:2025bmd,Xiong:2025obq,Ling:2025olz,Li:2020ejs,Li:2021gsx}. In this work, we formulate this reconstruction as a regularized inverse problem, using the available low-$|q^2|$ experimental constraints together with large-$|q^2|$ LCSR input. This approach mitigates part of the uncertainty associated with assuming a specific functional parametrization and leads to a complementary less parametrization-dependent determination of the transverse quark force in the proton. Ultimately, the results obtained in this work are consistent with those of Ref.\,\cite{Ji:2026lyj}, supporting that implied by a linear QCD potential, with manifestly reduced uncertainties in the small-$r_\perp$ region.

This paper is organized as follows. In Sec.\,\ref{sec:definition}, we introduce the definition of the force on quarks in the proton and its relation to the quark scalar form factor. In Sec.\,\ref{sec:reconstruction}, we present the improved reconstruction of the force on quarks using large-$|q^2|$ LCSR input and the inverse problem method, together with the resulting force distribution. We conclude with a summary and outlook in Sec.\,\ref{sec:summary}.

\section{Transverse color-Lorentz force in the IMF}
\label{sec:definition}

The force density acting on the quark sector is defined at the operator level by
\begin{equation}
 {\cal F}_q^i=\partial_\mu T_q^{\mu i}
 =g\,\bar\psi\gamma_\mu F^{\mu i}\psi \,,
\label{eq:force_operator}
\end{equation}
where the second equality follows from the QCD equations of motion.  The total quark-plus-gluon EMT is conserved; Eq.\,\eqref{eq:force_operator} describes the transfer of momentum between its quark and gluon parts. We use zero-skewness IMF kinematics,
\begin{equation}
 q^+=0 \,,\quad q^2=-\vec{q}_\perp^{\,2} \,,
 \quad q_\perp=|\vec{q}_\perp| \,,
\label{eq:kinematics}
\end{equation}
and denote the transverse impact parameter by $\vec{r}_\perp$, measured relative to the transverse center of momentum.  The distributions below are therefore two-dimensional light-front/impact-parameter distributions. They are not three-dimensional Breit-frame densities.

The nonconserved part of the quark EMT matrix element relevant to the force in the proton is written as
\begin{equation}
 \langle p'|T_q^{\mu\nu}|p\rangle_{\rm nc}
 =\bar u(p')\left[-\frac14 g^{\mu\nu}M G_{s,q}(q^2)\right]u(p) \,.
\label{eq:nc_emt}
\end{equation}
Here ``scalar form factor'' refers to the Lorentz-scalar coefficient in Eq.\,\eqref{eq:nc_emt}; it is not the conventional nucleon matrix element of $\bar q q$.  The relation between $G_{s,q}(q^2)$ and $A_q(q^2), B_q(q^2)$ and $C_q(q^2)$ has been given in Eq.\,\eqref{eq:Gsq_intro}.

We define the two-dimensional transform
\begin{align}
 \widetilde G_{s,q}(\vec{r}_\perp)
 &=\int\frac{d^2\vec{q}_\perp}{(2\pi)^2}
 e^{-i\vec{q}_\perp\cdot\vec{r}_\perp}
 G_{s,q}(q^2) \nn\\
 &=\frac{1}{2\pi}\int_0^\infty dq_\perp\,q_\perp
 J_0(q_\perp r_\perp)G_{s,q}(q^2) \,,
\label{eq:Gtilde}
\end{align}
where $r_\perp=|\vec{r}_\perp|$.  At leading order in the IMF expansion, the radial force density is
\begin{align}
 {\cal F}_q(\vec r_\perp)
 &=\frac{M}{4}\frac{d}{dr_\perp}\widetilde G_{s,q}(\vec r_\perp) \nn\\
 &=-\frac{M}{8\pi}\int_0^\infty dq_\perp\,q_\perp^2
 J_1(q_\perp r_\perp)G_{s,q}(q^2) \,.
\label{eq:force_density}
\end{align}

Following Ref.\,\cite{Ji:2026lyj}, an effective force per light quark is defined by dividing the force density by a transverse quark-number density,
\begin{equation}
 F_q(r_\perp)
 =\frac{\mathcal{F}_q(\vec r_\perp)}{\rho_q(\vec r_\perp)} \,.
\label{eq:force_per_quark}
\end{equation}
Using isospin symmetry and neglecting strange and heavier flavor corrections, the latter is approximated by
\begin{align}
 \rho_q(\vec{r}_\perp)&\simeq3\int\frac{d^2\vec{q}_\perp}{(2\pi)^2}
 e^{-i\vec{q}_\perp\cdot\vec{r}_\perp}
 \left[F_1^p(q^2)+F_1^n(q^2)\right] \nn\\
 &=\frac{3}{2\pi}\int_0^\infty dq_\perp\,q_\perp
 J_0(q_\perp r_\perp)\left[F_1^p(q^2)+F_1^n(q^2)\right] \,.
\label{eq:quark_density}
\end{align}
where $F_1^p(q^2)$ and $F_1^n(q^2)$ denote the electromagnetic Dirac form factors of the proton and neutron, respectively. They characterize the transverse distributions of electric charge in the IMF. The parametrizations of these form factors used in the numerical analysis are taken from Ref.\,\cite{Kelly:2004hm}. Combining Eq.\,\eqref{eq:force_density}--\eqref{eq:quark_density} gives
\begin{equation}
 F_q(r_\perp)=-\frac{M}{12}
 \frac{\displaystyle\int_0^\infty dq_\perp\,q_\perp^2
 J_1(q_\perp r_\perp)G_{s,q}(q^2)}
 {\displaystyle\int_0^\infty dq_\perp\,q_\perp
 J_0(q_\perp r_\perp)\left[F_1^p(q^2)+F_1^n(q^2)\right]} \,.
\label{eq:force}
\end{equation}
In natural units, $F_q(r_\perp)$ has mass dimension two. A negative radial component denotes an inward attractive force.

\section{Improved Reconstruction of the Force on Quarks}
\label{sec:reconstruction}
In this section, we improve the theoretical framework proposed in Ref.\,\cite{Ji:2026lyj}: First, we supplement the low-$|q^2|$ information with LCSR input at larger $|q^2|$; second, we reconstruct the radial force density directly as a regularized inverse problem.

The low-$|q^2|$ constraints are taken from the phenomenological DVCS analysis, which extracted the so called quark $D$-term form factor through a dispersion relation analysis \cite{Burkert:2018bqq}, and from the GUMP analysis of Ref.\,\cite{Guo:2025muf}, which performed an NLO global extraction of nucleon GPDs by combining hard exclusive data with PDF, electromagnetic form factor, and lattice QCD inputs. These constraints are therefore indirect extractions of the relevant EMT FFs rather than pointwise measurements of $G_{s,q}(q^2)$. At larger $|q|^2$, we use the flavor-decomposed LCSR calculation of Ref.\,\cite{Dehghan:2025ncw}, which evaluates valence-quark EMT FFs while omitting explicit gluon contributions.  Fig.\,\ref{fig:input-points} shows the representative scalar form factor points used in the numerical illustration: six low-$|q^2|$ constraints in $0\leq -q^2\leq0.35~{\rm GeV}^2$ and 20 sampled LCSR points in $1\leq -q^2\leq10~{\rm GeV}^2$. At present, we assign a $10\%$ relative uncertainty to each point; a more detailed discussion of the LCSR uncertainties is provided below.

The large-$|q^2|$ input is taken from the LCSR calculation of the proton EMT FFs in Ref.\,\cite{Dehghan:2025ncw}. LCSR is a well-established nonperturbative method for hadronic form factors at intermediate and large spacelike momentum transfer. Its reliability in this region follows from the light-cone operator product expansion, where the short-distance hard kernel is factorized from universal hadron distribution amplitudes, while the Borel transformation and continuum subtraction suppress excited-state contributions \cite{Balitsky:1989ry,Colangelo:2000dp,Azizi:2019ytx}. This makes LCSR complementary to the low-$|q^2|$ experimental and lattice information and particularly useful for the Fourier-Bessel reconstruction of short and intermediate transverse distances.

Eq.\,(\ref{eq:force_density}) is the first-order Hankel transform of $G_{s,q}(q^2)$. In practice, only a finite set of discrete values of $G_{s,q}(q^2)$ is available, rather than its complete momentum space dependence. Fitting these data with a prescribed functional form would inevitably introduce dependence on the chosen parametrization, particularly in momentum regions where the available constraints are sparse. An alternative is to formulate the reconstruction as an inverse problem. By rewriting the Hankel-transform relation as an integral equation in which the discrete values of $G_{s,q}(q^2)$ constitute the input and $\mathcal{F}_q(\vec r_\perp)$ is the unknown function, one can reconstruct $\mathcal{F}_q(\vec r_\perp)$ directly without first assuming a specific parametrization of $G_{s,q}(q^2)$. To this end, we define the corresponding inverse transform
\begin{equation}
    G_{s,q}(q^2) =  -\frac{8\pi}{M q_\perp}\int_0^\infty dr_\perp\, r_\perp\, J_1(q_\perp r_\perp)\, \mathcal{F}_q(\vec r_\perp)\, ,
    \label{Inverse_Problem_Integral_Equation}
\end{equation}
with $q_\perp > 0$. The smooth limit $J_1(q_\perp r_\perp)/q_\perp \to r_\perp/2$ as $q_\perp \to 0$ yields the zero-momentum value
\begin{equation}
    G_{s,q}(0) =  -\frac{4\pi}{M}\int_0^\infty dr_\perp\, r_\perp^2\, f_q(r_\perp)\,.
    \label{eq:zeroMomentumRow}
\end{equation}
Importantly, none of the data points determines $\mathcal{F}_q(\vec r_\perp)$ at a particular transverse distance. Instead, each point of $G_{s,q}(q^2)$ constrains a weighted integral of $\mathcal{F}_q(\vec r_\perp)$ over the entire spatial domain. Reconstructing the full function from a finite and uncertain set of such integral constraints is therefore an inverse problem. In this formulation, the unknown is not a parametrized multipole or dipole form of $G_{s,q}(q^2)$ but the function $\mathcal{F}_q(\vec r_\perp)$ itself. The objective is to extract this function directly from Eqs.\,\eqref{Inverse_Problem_Integral_Equation} and \eqref{eq:zeroMomentumRow}, without assuming a specific functional dependence for $G_{s,q}(q^2)$.

For the numerical inversion we truncate the $r_\perp$ integral at $r_{\perp \rm max}=10\,{\rm GeV}^{-1}$ and use a uniform grid $r_{\perp j}=j\Delta r_\perp$. With the trapezoidal weights $w_j$, Eqs.\,(\ref{Inverse_Problem_Integral_Equation}) and (\ref{eq:zeroMomentumRow}) become the linear system
\begin{equation}
    \begin{aligned}
        G_i = A_{i j}\, f_j \,,
    \end{aligned} \label{Inverse_Problem_Matrix}
\end{equation}
where $f_j=\mathcal{F}_q(r_{\perp j})$, $G_i$ is the input-data vector and
\begin{equation}
 A_{ij}=
 \begin{cases}
 \displaystyle -\frac{8\pi }{Mq_{\perp i}}r_{\perp j} w_j J_1\left(q_{\perp i} r_{\perp j}\right) \,, &q_{\perp i}>0 \,,\\[6pt]
 \displaystyle -\frac{4\pi}{M} r_{\perp j}^2w_j \,, &q_{\perp i}=0 \,.
 \end{cases}
\label{eq:matrix_kernel}
\end{equation}
Here $i$ indexes the measured $q_\perp^2$  points and $j$ is the node index in the $r_\perp$-space discretization. The singular value decomposition reveals a condition number for $A$ of order $10^{16}$, showing that the matrix is severely ill-conditioned~\cite{allen1985singular}. As a result, any small, unavoidable errors in the input $G_i$ are dramatically amplified in the solution $f_j$, precluding a direct inversion of Eq.\,(\ref{Inverse_Problem_Matrix}). To  stabilize the inversion, we adopt Tikhonov regularization, which imposes a smoothness constraint on the solution~\cite{Xiong:2022uwj,Xiong:2025bmd,Xiong:2025obq,Ling:2025olz}.

In the Tikhonov approach, one minimizes the Tikhonov functional
\begin{equation}
    \begin{aligned}
        f_\alpha=\arg \min \left\{ \left\|A f - G\right\|_2^2 + \alpha  \left\|L f\right\|_2^2 \right\}\,.
    \end{aligned}\label{Tik_regularization}
\end{equation}
and the minimizer equivalently satisfies the normal equation
\begin{equation}
\left(A^TA+\alpha L^TL\right)f_\alpha=A^TG \,,
\label{eq:tikhonovNormal}
\end{equation}
where $\| \cdot \|_2$ denotes the Euclidean norm, the matrix $L$ encodes prior information about the solution, and $\alpha>0$ is the regularization parameter. 
The first term in Eq.\,(\ref{Tik_regularization}) enforces fidelity to the data. Because the physical force distribution is expected to be smooth, the second term suppresses spurious oscillations through a high-order derivative matrix $L$. We choose $L$ to be the third-derivative operator, which penalizes rapid, nonphysical fluctuations while leaving slowly varying features largely unaffected. Higher-order derivatives offer no substantial advantage and are not adopted here. Discretizing $L$ with the third-order finite-difference formula,
\begin{equation}
\left(Lf \right)_j=\frac{-f_j+3f_{j+1}-3f_{j+2}+f_{j+3}}{\left(\Delta r_\perp\right)^3} \,,
\label{eq:L3operator}
\end{equation}
efficiently filters out high-frequency noise while preserving the smooth force distribution. The method is reliable for the present purpose because it is a controlled regularization of an explicitly known linear inverse problem, the regularization strength is selected from the data, and similar inverse-transform strategies have been tested in nonperturbative QCD applications such as quasi-distribution reconstructions in LaMET \cite{Xiong:2025obq,Ling:2025olz}.

\begin{figure}[t]
    \centering
    \includegraphics[width=\linewidth]{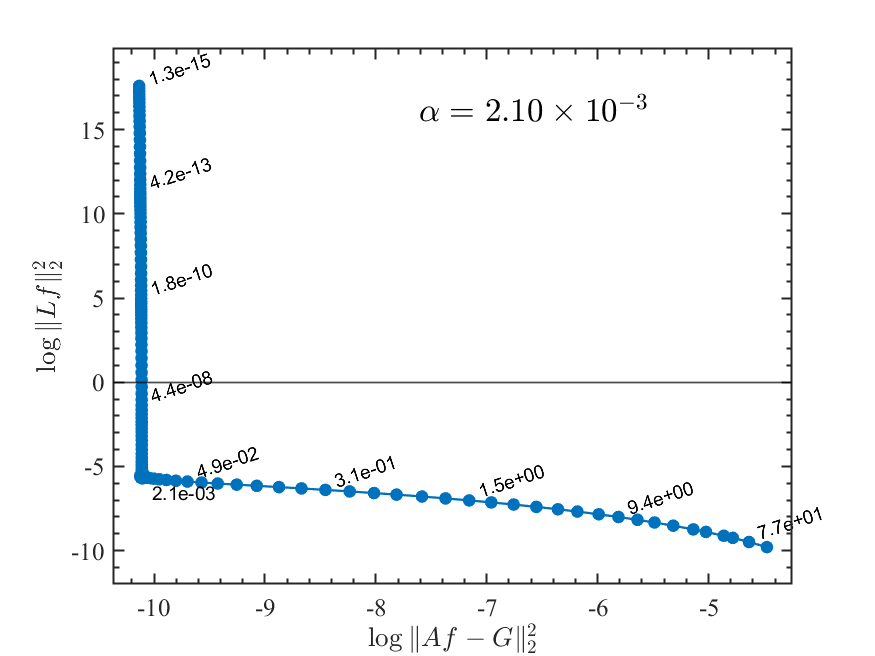}
    \caption{L-curve used to determine the Tikhonov regularization parameter. The selected point of maximum curvature corresponds to $\alpha=2.1\times10^{-3}$.}
    \label{fig:l-curve}
\end{figure}

We choose $\alpha$ with the L-curve method, which identifies the point of maximum curvature on the log-log plot of the residual norm $\|A f_\alpha - G\|_2$ versus the penalty norm $\|L f_\alpha\|_2$. As $\alpha \to 0$, the residual term dominates, signaling overfitting through noise amplification driven by the ill-conditioning of $A$; as $\alpha \to \infty$, the regularization term dominates, leading to underfitting and oversmoothing of the solution. The corner of the L-curve corresponds to a compromise value of $\alpha$ that balances data fidelity and smoothness. In this method, the regularization parameter is determined systematically from the data, thus further reducing parametrization dependence. As shown in Fig.\,\ref{fig:l-curve}, the point of maximum curvature yields $\alpha=2.1\times10^{-3}$.

\begin{figure}[t]
    \centering
    \includegraphics[width=0.9\linewidth]{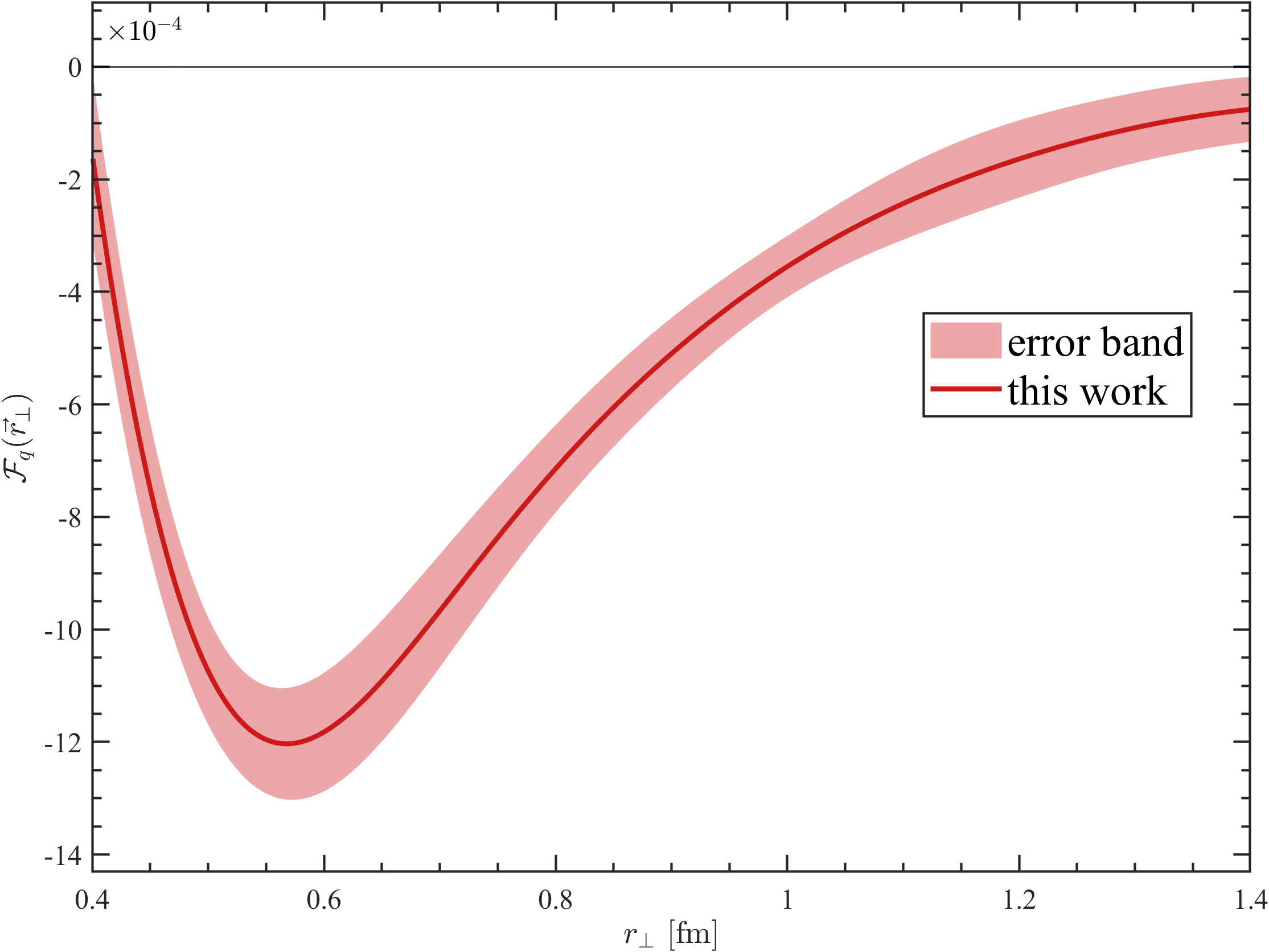}
    \caption{The resulting $\mathcal{F}_q(\vec r_\perp)$ obtained from the inverse problem method. It is derived from 26 input points with their associated errors presented in Fig.\,\ref{fig:input-points}.}
    \label{fig:force-density}
\end{figure}

Through the inverse problem reconstruction, we obtain the function $\mathcal{F}_q(\vec r_\perp)$ defined in Eq.\,\eqref{eq:force_density}. Its numerical result is shown in Fig.\,\ref{fig:force-density}, which exhibits a smooth and stable profile without evident unphysical oscillations, indicating that the inverse procedure yields a well-behaved solution. It is worth noting that, in the interval $1.1\leq r_\perp\leq1.4\,\mathrm{fm}$, the magnitude of the central value of $\mathcal{F}_q(\vec r_\perp)$ decreases with its absolute uncertainty remaining nearly unchanged. The advantage of the present analysis is twofold. First, the inclusion of LCSR information at $1\leq |q^2| \leq10\,{\rm GeV}^2$ supplies direct constraints on the momentum region to which small and intermediate $r_\perp$ are most sensitive, so the uncertainty in this transverse distance region is visibly reduced. Second, once the discrete scalar form factor points in Fig.\,\ref{fig:input-points} are supplied, no specific assumption of the $q^2$ dependence of $A_q(q^2)$, $B_q(q^2)$, $C_q(q^2)$, or $G_{s,q}(q^2)$ is needed. The force is obtained by solving the regularized inverse problem itself, making the result a less parametrization-dependent confirmation of the confinement force signal found in Ref.\,\cite{Ji:2026lyj}.

\begin{figure*}
    \centering
    \includegraphics[width=0.82\textwidth]{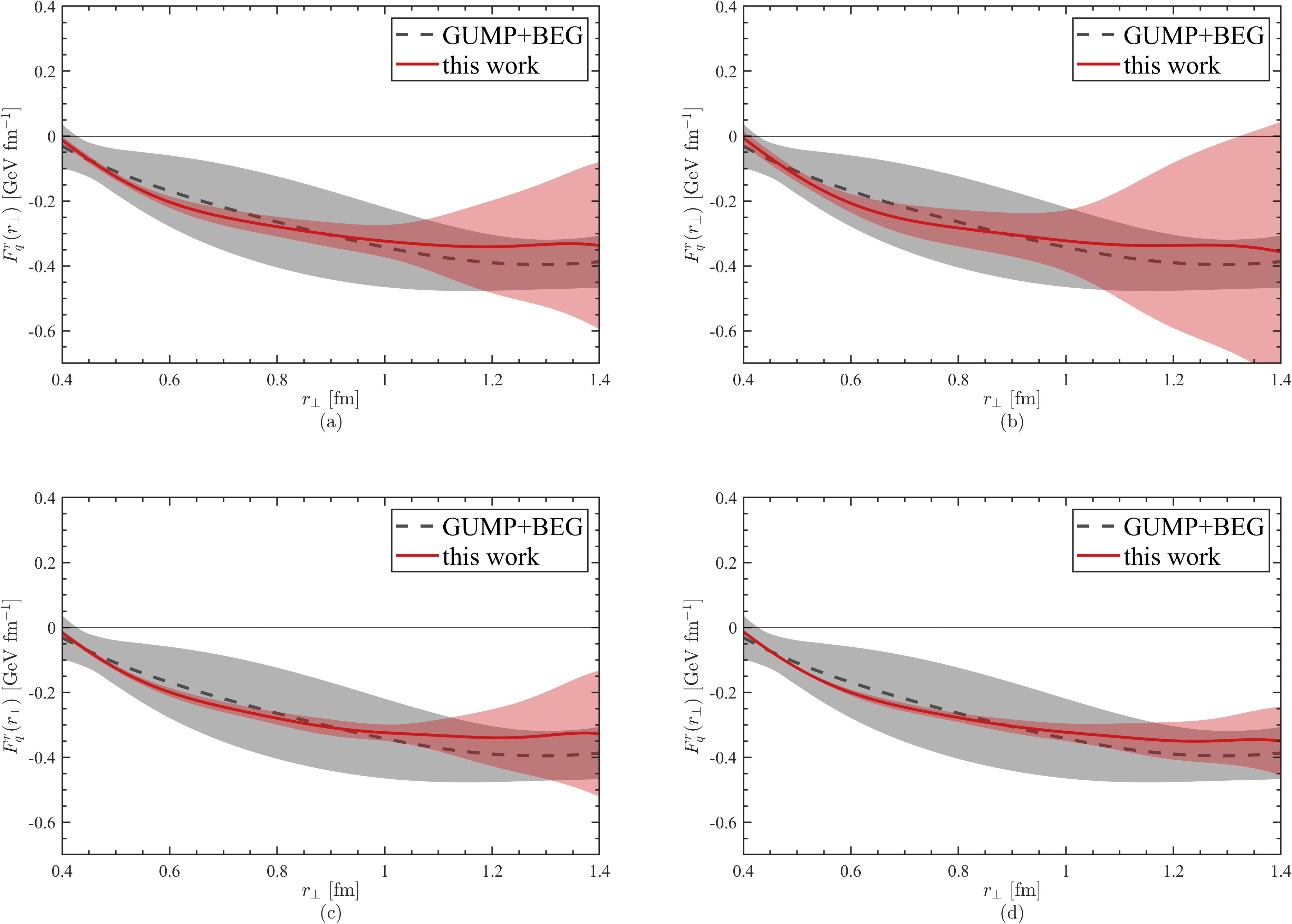}
    \caption{Quark force reconstructed using inverse problem method with the L-curve choice of the regularization parameter. The black curves show the results of Ref.\,\cite{Ji:2026lyj}. The red curves in panels (a) and (b) are obtained using the  GUMP \cite{Guo:2025muf} and BEG \cite{Burkert:2018bqq} data supplemented with LCSR \cite{Dehghan:2025ncw} inputs, to which relative uncertainties of $10\%$ and $20\%$, respectively, are assigned. The red curve in panel (c) is obtained using the GUMP and BEG data supplemented with LCSR inputs assigned a $5\%$ relative uncertainty. The red curve in panel (d) is obtained by additionally including two synthetic data points at $-q^2=0.6$ and $0.8\,\mathrm{GeV}^2$.}
    \label{fig:force-band}
\end{figure*}

By dividing the force density by the transverse quark-number density, as indicated in Eq.\,(\ref{eq:force_per_quark}), we obtain the final force acting on the quarks in the proton, which is presented in Fig.\,\ref{fig:force-band}. Negative value corresponds to an inward attractive force. In the intermediate region the force is approximately constant and attractive; for example, the average over $0.7\leq r_\perp\leq 1.1\,{\rm fm}$ is about $-0.3\,{\rm GeV/fm}$. This agrees, within uncertainties and conventions, with the result of Ref.\,\cite{Ji:2026lyj}, and therefore supports the same physical conclusion: the resulting force is consistent with that implied by a linear QCD potential in the corresponding transverse distance region.

As seen in Fig.\,\ref{fig:force-band}, the uncertainty of the quark force derived in this work is considerably reduced compared to that in Ref.\,\cite{Ji:2026lyj} for $0.5 \leq r_\perp \leq \text{1.0\,fm}$. This improvement stems from the incorporation of LCSR calculations at intermediate and large-$|q^2|$. Although $r_\perp$ denotes a transverse impact parameter rather than a three-dimensional radial coordinate, this interval covers the characteristic proton size indicated by its charge radius of approximately $0.84\,\mathrm{fm}$. The markedly reduced uncertainty therefore establishes the attractive character of the force with greater confidence, providing stronger force-based evidence for quark confinement in the proton. The $10\%$ relative uncertainty assigned to the LCSR inputs is adopted as an exploratory benchmark, since the available LCSR calculation only estimates uncertainties through variations of its auxiliary parameters and nonperturbative inputs \cite{Dehghan:2025ncw}. Our result enables a transparent assessment of the constraining power of the large-$|q^2|$ information, while the impact of improved input precision will be examined below. 

For $r_\perp>\text{1.1\,fm}$, the uncertainty of the inverse problem solution increases rapidly. As we have discussed in Fig.\,\ref{fig:force-density}, when $r_\perp$ increases from $1.1$ to $1.4\,\mathrm{fm}$, the magnitude of the central value of $\mathcal{F}_q(\vec r_\perp)$ decreases by approximately a factor of three, whereas its absolute uncertainty remains nearly unchanged. Consequently, the relative uncertainty increases by approximately the same factor. After division by the quark-number density, the central value of the resulting force approaches a constant over this interval, while its relative uncertainty still increases by approximately a factor of three. This explains the rapid growth of the uncertainty beyond $r_\perp=1.1\,\mathrm{fm}$ observed in Fig.\,\ref{fig:force-band}. One can readily observe that, with a $20\%$ uncertainty assigned to the LCSR input, the error bands in Fig.\,\ref{fig:force-band}(b) already extend above zero at large $r_\perp$. But it is still worth noting that, even with a $20\%$ uncertainty assigned to the LCSR inputs, the uncertainty of the force remains well controlled in the interval $0.5\leq r_\perp\leq 1.0\,\mathrm{fm}$, where the force retains a clearly attractive character. The uncertainties in Ref.\,\cite{Ji:2026lyj} are smaller than ours in large-$r_\perp$ region. Since they employ a parametric form for the form factors $A_q(q^2)$ and $C_q(q^2)$, thus the quoted errors do not include the parametrization dependence.

These observations show that, although the inverse problem reconstruction avoids assuming a specific functional parametrization, the precision of the resulting force remains sensitive to the accuracy of the input information, including experimental extractions, lattice QCD simulations, and LCSR calculations. To assess the level of input precision required to draw a firmer conclusion about quark confinement, we perform a prospective test in which the relative uncertainty assigned to the LCSR input is reduced to $5\%$. The resulting force distribution is shown in Fig.\,\ref{fig:force-band}(c). The uncertainty is substantially reduced over the entire $r_\perp$ region; nevertheless, it still grows rapidly for $r_\perp\gtrsim1.1\,\mathrm{fm}$.


From an inverse-problem perspective, the persistent uncertainty at large $r_\perp$ indicates missing kinematic coverage rather than merely insufficient pointwise precision. Fig.\,\ref{fig:input-points} reveals a gap in the input coverage over $0.35\leq -q^2\leq1.0\,\mathrm{GeV}^2$. Following the parametrization of Ref.\,\cite{Ji:2026lyj}, we therefore introduce two synthetic constraints in this transition region,
\begin{equation*}
\begin{aligned}
G_{s,q}(q^2)&=0.000\pm 0.002 &&\text{at}&& -q^2=0.6\,\text{GeV}^2\,,\\
G_{s,q}(q^2)&=-0.040\pm 0.002 &&\text{at}&& -q^2=0.8\,\text{GeV}^2\,,
\end{aligned}
\end{equation*}
and assign a $5\%$ relative uncertainty to each of them. The corresponding result is displayed in Fig.\,\ref{fig:force-band}(d), where the uncertainty remains small throughout the full $r_\perp$ region. These tests identify two complementary priorities for future studies: extending the input information toward large-$|q^2|$ and obtaining further constraints in the transition region $0.35\leq -q^2\leq1.0\,\mathrm{GeV}^2$. More precise and complete inputs in these regions will enable a more tightly constrained determination of the force on quarks and help clarify the dynamical origin of quark confinement.

\section{Summary and Outlook}
\label{sec:summary}

Confinement remains one of the most important issues in particle physics, and the recent work of Ref.\,\cite{Ji:2026lyj} has provided a new perspective for understanding it directly from QCD through the force acting on quarks in the proton. In this work, we revisited this quark force construction by supplementing the limited experimental and lattice QCD information at large spacelike momentum transfer with LCSR results for the proton EMT FFs. We also formulated the determination of the transverse quark force $F_q(r_\perp)$ as a regularized inverse problem. The resulting force distribution provides a complementary and less parametrization-dependent determination of the confinement force inside the proton. It remains consistent with that implied by a linear QCD potential and with the attractive, approximately constant force observed in Ref.\,\cite{Ji:2026lyj}, while the inclusion of large-$|q^2|$ information improves the reconstruction in the short-distance region. We also show how advances in future experimental, lattice QCD, and LCSR studies can significantly improve the precision of the reconstructed quark force.

The present analysis shows how to directly determine the strong force on quarks within the proton and provides a useful tool for studying confinement from a force-distribution perspective. Nevertheless, this work should be regarded as an initial attempt, and several further developments are required before a quantitatively conclusive determination can be reached. First, the transverse quark-number density $\rho_q(\vec r_\perp)$, which enters the denominator of the force in Eq.\,(\ref{eq:force_per_quark}), is itself reconstructed from the electromagnetic Dirac form factors, and its uncertainties should be consistently propagated. Second, a systematic comparison of different regularization schemes is needed to assess the methodological dependence of the reconstructed force. Third, as mentioned above, the current LCSR results retain dependence on model and auxiliary parameters; in future work, we plan to formulate the determination of the EMT FFs themselves as an inverse problem, thereby reducing this dependence. Accurate inputs in the currently less constrained interval $0.35\leq -q^2\leq1.0\,\mathrm{GeV}^2$ will also be essential for controlling the force uncertainty. This framework can be naturally extended to quark confinement in mesons, confinement mechanisms in exotic hadrons, and force distributions associated with gluons inside hadrons. Together with future data from electron-ion colliders and other experiments, such theoretical developments may eventually lead to a more complete picture of how partons are bound into hadrons by the strong interaction.

\section*{Acknowledgements}
This work is supported in part by the National Natural Science Foundation of China under Grant No. 12335003, and by the Fundamental Research Funds for the Central Universities under No. lzujbky-2023-stlt01, lzujbky-2024-oy02 and lzujbky-2025-eyt01, the Scientific Research Innovation Capability Support Project for Young Faculty under Grant No. ZYGXQNJSKYCXNLZCXM-P2. J.X is supported in part by the National Natural Science Foundation of China under Grant No. 12475098, 12105247, and by the Key Laboratory for Particle Astrophysics and Cosmology, Ministry of Education (MoE).

\end{document}